\documentclass[11pt]{article}
\usepackage{amssymb}
\makeatletter\@addtoreset {equation}{section}\makeatother

\newtheorem{theorem}{Theorem}[section]
\newtheorem{lemma}[theorem]{Lemma}

\newtheorem{proposition}[theorem]{Proposition}

\newtheorem{assumption}[theorem]{Assumption}

\newtheorem{remark}[theorem]{Remark}
\newenvironment{proof}{
    \noindent {\it Proof.}}{\hfill$\Box$\\ \medskip
}

 {\begin{trivlist} \item[]{\bf Example}}%
 {\hspace*{\fill}$\rule{.3\baselineskip}{.35\baselineskip}$\end{trivlist}}

\usepackage[dvips]{epsfig}
\usepackage{graphicx}

\begin{document}

\title{\bf Spectral stability and time evolution \\ of N-solitons in KdV hierarchy}
\author{Yuji Kodama$^{\dagger}$ and Dmitry Pelinovsky$^{\dagger \dagger}$ \\
$^{\dagger}$ Department of Mathematics, Ohio State University,
Columbus, OH 43210, USA \\
$^{\dagger \dagger}$ Department of Mathematics, McMaster University,
Hamilton, Ontario, Canada, L8S 4K1 }
\date{\today}
\maketitle

\begin{abstract}
This paper concerns spectral stability and time evolution of $N$-solitons
in the KdV hierarchy with mixed commuting time flows. Spectral
stability problem is analyzed by using a pair of self-adjoint
operators with finite numbers of negative eigenvalues. We show that
the absence of unstable eigenvalues in the stability problem is
related to the absence of negative eigenvalues of these operators in
the constrained function spaces. Time evolution of $N$-solitons is
uniquely characterized from the inverse scattering transform
technique.
\end{abstract}

\section{Introduction}

We address the stability problem for solitary waves in the KdV-type
evolution equation:
\begin{equation}
\label{generalKdV} \frac{d u}{d t} = \frac{\partial}{\partial x}
\frac{\delta H}{\delta u},
\end{equation}
where $u \in \mathbb{R}$, $x \in \mathbb{R}$, $t \in \mathbb{R}_+$,
and $H(u)$ is the Hamiltonian. Besides the canonical Korteweg--de
Vries (KdV) equation \cite{Newell}, the time-evolution problem
(\ref{generalKdV}) includes higher-order KdV equations of the
integrable KdV hierarchy (e.g. see \cite{KN78,KT78}). The integrable
KdV hierarchy can be derived with the asymptotic multi-scale
expansion technique for modelling of solitary waves in physical
non-integrable problems, e.g. in ferromagnets
\cite{Leblond1,Leblond2} (see also \cite{HK02} for the normal form of perturbed
KdV equation).

Our work originates from analysis of spectral stability of solitary
waves in the KdV-type evolution equations
\cite{BSS87,Wen,SS90,PW92}. Let us assume that the evolution problem
(\ref{generalKdV}) has a solitary wave solution $u = u_0(x,t)$ that
decays exponentially in space $x \in \mathbb{R}$ and changes in time
$t \in \mathbb{R}_+$ according to symmetries of (\ref{generalKdV}).
The spectral stability problem for KdV solitary waves takes the
general form:
\begin{equation}
\label{spectrum}
\partial_x {\cal L} v = \lambda v,
\end{equation}
where $v \in \mathbb{C}$, $\lambda \in \mathbb{C}$, and the
self-adjoint linearized operator ${\cal L}$ is computed at the
solution $u_0(x,t)$ after separation of variables $(x,t)$, i.e. ${\mathcal L}=D^2H(u_0)$. The
solitary wave solution $u_0(x,t)$ is spectrally unstable if there
exist an eigenvalue $\lambda \in \mathbb{C}$ with ${\rm
Re}(\lambda) > 0$ and $v \in L^2(\mathbb{R})$. It is weakly
spectrally stable if no eigenvalues $\lambda \in \mathbb{C}$ with
${\rm Re}(\lambda) > 0$ and $v \in L^2(\mathbb{R})$ exist.

We shall characterize unstable eigenvalues of the stability problem
(\ref{spectrum}) from the study of the self-adjoint operator ${\cal
L}$. The operator ${\cal L}$ defines the energy quadratic form:
\begin{equation}
\label{energy-quadratic-form} h(v) = \left( v, {\cal L} v \right):=\int_{\mathbb R}v(x)({\mathcal L}v)(x)\,dx.
\end{equation}
The relation between eigenvalues of the stability problem and those
of the linearized energy was recently studied in the context of
spectral stability of solitary waves in the nonlinear
Schr\"{o}dinger equations \cite{CPV03,P03}. The number of unstable
eigenvalues in the spectral stability problem was found to be
related to the number of negative eigenvalues of the energy
quadratic form in a constrained function space. The same relation
was derived independently in \cite{KKS03} by extending earlier
results \cite{G90} to an abstract Hamiltonian dynamical system.

The methods of \cite{KKS03,P03} are not applicable to the KdV-type
evolution equations, since the symplectic operator $\partial_x$ is
not invertible. Nevertheless, we will show with explicit analysis of
quadratic forms that the spectral stability problem (\ref{spectrum})
can be embedded into a larger problem, which has the same structure
as that considered in \cite{KKS03,P03}. Since the operator ${\cal
L}$ maps $L^2(\mathbb{R})$ to $L^2(\mathbb{R})$, the eigenfunction
$v(x) \in L^2(\mathbb{R})$ for $\lambda \neq 0$ must satisfy the
constraint:
\begin{equation}
\label{mass-constraint} (1, v) = \int_{\mathbb{R}} v(x) dx = 0.
\end{equation}
If $v \in L^2(\mathbb{R})$ decays exponentially as $|x| \to \infty$
and satisfies the constraint (\ref{mass-constraint}), the
eigenfunction of the spectral problem (\ref{spectrum}) can be
represented as $v = - w'(x)$, where $w \in L^2(\mathbb{R})$, i.e.
$v\in {\rm im}(\partial_x)$. We
shall hence replace the stability problem (\ref{spectrum}) by the
coupled system:
\begin{equation}
\label{coupled-problem} {\cal L} v = - \lambda w, \qquad {\cal M} w
= \lambda v,
\end{equation}
where ${\cal M} = - \partial_x {\cal L} \partial_x$. Eliminating $w$
from the system (\ref{coupled-problem}), we reduce the coupled
problem (\ref{coupled-problem}) to the scalar problem:
\begin{equation}
\label{squared-problem}
\partial_x {\cal L} \partial_x {\cal L} v = \lambda^2 v,
\end{equation}
which has both eigenvalues $\lambda$ and $-\lambda$ of the original
problem (\ref{spectrum}). Moreover, if $\lambda$ is a simple
eigenvalue of $\partial_x {\cal L}$, then the system
(\ref{coupled-problem}) has the solution $v = -w'(x)$, while if
$\lambda$ is a simple eigenvalue of $-\partial_x {\cal L}$, then the
system (\ref{coupled-problem}) has the solution $v = w'(x)$,

The coupled system (\ref{coupled-problem}) is defined in a
constrained subspace of $L^2(\mathbb{R},\mathbb{C}^2)$. Assuming
that the linear operator ${\cal L}$ has an isolated kernel and a
positive continuous spectrum and that the eigenfunctions of ${\rm
ker}({\cal L})$ satisfy the constraint (\ref{mass-constraint}), we
introduce the constrained subspaces:
\begin{eqnarray}
\label{Xc} X_c(\mathbb{R}) & = & \left\{ v \in L^2(\mathbb{R}) :
\; (v, {\rm ker}({\cal L} \partial_x) ) = 0 \right\}, \\
\label{Xc'} X_c'(\mathbb{R}) & = & \left\{ w \in L^2(\mathbb{R}) :
\; (w, {\rm ker}({\cal L}) ) = 0 \right\}.
\end{eqnarray}
If $(v,w) \in L^2(\mathbb{R},\mathbb{C}^2)$ is the eigenvector of
the system (\ref{coupled-problem}) for $\lambda \neq 0$, then $v
\in X_c(\mathbb{R})$ and $w \in X_c'(\mathbb{R})$.

Within this general formalism, we shall study $N$-solitons of the
KdV hierarchy. Time evolution of the $N$-solitons is defined by the
mixed commuting flows of the higher-order KdV equations. Using
earlier results of \cite{MS93}, we shall prove that the operators
${\cal L}$ and ${\cal M}$ have no negative eigenvalues in the
constrained spaces $X_c(\mathbb{R})$ and $X_c'(\mathbb{R})$,
respectively. This fact has a consequence that the stability problem
(\ref{spectrum}) and the coupled system (\ref{coupled-problem}) have
no unstable eigenvalues with ${\rm Re}(\lambda) > 0$. We also use
the inverse scattering transform technique \cite{AK82} to
characterize the time evolution of $N$-solitons in the mixed
commuting flows of the KdV hierarchy. Uniqueness of $N$-solitons in
a constrained variational problem is also proved. Both main results
(spectral stability and time evolution) are not covered by the
previous publications on $N$-solitons in the KdV hierarchy
\cite{MS93,MMT02}.

The paper is organized as follows. In Section 2, we briefly summarize the background of the integrable KdV hierarchy, and give an explicit definition of the spaces $X_c({\mathbb R})$ and $X_c'({\mathbb R})$
for the $N$-solitons. Spectral stability of $N$-solitons in mixed
commuting time flows is studied in Section 3. Time evolution of
$N$-solitons is characterized in Section 4. Section 5 concludes the
paper.

\section{Review of $N$-solitons in the KdV hierarchy}

The integrable KdV hierarchy (see review in \cite{Newell}) is
defined by the set of KdV-type evolution equations:
\begin{equation}
\label{KdVhierarchy} \frac{\partial u}{\partial t_n} =
\frac{\partial}{\partial x} \frac{\delta H_n}{\delta u}, \qquad n
\in \mathbb{N}_+,
\end{equation}
where $H_n(u)$ are the Hamiltonian in energy space $H^{n-1}({\Bbb
R})$. The Hamiltonians are constructed recursively as
\begin{equation}
\label{recursion} {\cal J} \frac{\delta H_{n+1}}{\delta u} = {\cal
K} \frac{\delta H_n}{\delta u}, \quad n \in \mathbb{N}_+,
\end{equation}
where the linear operators ${\cal J}$ and ${\cal K}$ take the form:
\begin{equation}
\label{recursion_operators} {\cal J} = \frac{\partial}{\partial x},
\qquad {\cal K} = \frac{\partial^3}{\partial x^3} + 2 \left(
\frac{\partial}{\partial x} u + u \frac{\partial}{\partial x}
\right)
\end{equation}
and the lowest three Hamiltonians are:
\begin{eqnarray}
H_1 & = & \frac{1}{2} \int_{\mathbb{R}} u^2 dx, \\
H_2 & = & \frac{1}{2} \int_{\mathbb{R}} \left( u_x^2 - 2 u^3 \right)
dx, \\ \label{HamiltonianH3} H_3 & = & \frac{1}{2} \int_{\mathbb{R}}
\left( u_{xx}^2 - 10 u u_x^2 + 5 u^4 \right) dx.
\end{eqnarray}
The Hamiltonians $H_1$, $H_2$, and $H_3$ generate the first three
members of the KdV hierarchy, which are given by the transport
equation $u_{t_1} = u_x$, the KdV equation,
\begin{equation}
\label{KdV} u_{t_2} = - u_{xxx} - 6 u u_x,
\end{equation}
and the integrable fifth-order KdV equation,
\begin{equation}
\label{fifthKdV} u_{t_3} = u_{xxxxx} + 10 u u_{xxx} + 20 u_x u_{xx}
+ 30 u^2 u_x.
\end{equation}
All members of the KdV hierachy
(\ref{KdVhierarchy})--(\ref{HamiltonianH3}) have families of
$(2N)$-parameter solutions called $N$-solitons which are expressed
by the $\tau$-function \cite{Newell},
\begin{equation}
\label{Nsoliton} u(x;t_1,t_2,...)  = U_N(x-\theta_1,...,x-\theta_N)
= 2 \frac{\partial^2}{\partial x^2} \log \tau(x - \theta_1,...,x -
\theta_N),
\end{equation}
where
\begin{equation}
\label{theta} \theta_n(t_1,t_2,...) = \delta_n + \sum_{k=1}^{\infty}
(-1)^k c_n^{k-1} t_k = \delta_n - t_1 + c_n t_2 - c_n^2 t_3 + ...,
\end{equation}
and $n = 1,2,...,N$. Parameters $\delta_n$ are arbitrary, while
parameters $c_n$ satisfy the constraints $c_n > 0$ for all $n =
1,2,...,N$.

We give explicit formulae for the first two solitons
(\ref{Nsoliton})--(\ref{theta}). The $1$-soliton is given by:
\begin{equation}
\label{1-sol} U_1 = 2 \frac{\partial^2}{\partial x^2} \log\left[ 1 +
e^{\sqrt{c_1}(x-\theta_1)} \right], \qquad c_1 > 0.
\end{equation}
The function $U_1(x)$ is a decaying solution of the second-order
ODE:
\begin{equation}
\label{1-sol-ODE} H_2'(u) + \nu_1 H_1'(u) = - u_{xx} - 3 u^2 + c_1 u
= 0,
\end{equation}
where $\nu_1 = c_1$. The $2$-solitons are given by:
\begin{equation}
\label{2-sol} U_2 = 2 \frac{\partial^2}{\partial x^2} \log\left[ 1 +
e^{\sqrt{c_1}(x-\theta_1)} + e^{\sqrt{c_2}(x-\theta_2)} + \left(
\frac{\sqrt{c_1} - \sqrt{c_2}}{\sqrt{c_1} + \sqrt{c_2}}\right)^2
e^{\sqrt{c_1}(x-\theta_1) + \sqrt{c_2}(x-\theta_2)}\right],
\end{equation}
where $c_1, c_2 > 0$ and $c_1 \neq c_2$. The function $U_2(x)$ is a
decaying solution of the fourth-order ODE:
\begin{eqnarray}
\nonumber H_3'(u) + \nu_2 H_2'(u) + \nu_1 H_1'(u) & = & \\
\label{2-sol-ODE} u_{xxxx} + 10 u u_{xx} + 5 u_x^2 + 10 u^3 - (c_1 +
c_2) (u_{xx} + 3 u^2) + c_1 c_2 u & = & 0,
\end{eqnarray}
where $\nu_1 = c_1 c_2$ and $\nu_2 = c_1 + c_2$.

In general, the functions $U_N(x)$ are critical points of the
Lyapunov functional in $H^N({\Bbb R})$:
\begin{equation}
\label{Lyapunov} \Lambda_N(u) = H_{N+1}(u) + \sum_{n=1}^N \nu_n
H_n(u),
\end{equation}
such that
\begin{equation}
\label{Lyapunov-variation} \Lambda_N'(U_N) = H_{N+1}'(U_N) +
\sum_{n=1}^N \nu_n H_n'(U_N) = 0,
\end{equation}
where the Lagrange multipliers $\nu_1$,...,$\nu_N$ are elementary
symmetric functions of $c_1$,...,$c_N$ due to normalization of $H_n$
(see \cite{MS93}). Lyapunov stability of $N$-solitons as critical
points of the constrained Hamiltonian $H_{N+1}(u)$ was proved in
\cite{MS93}. Specifically, $N$-solitons $U_N(x)$ are minimal points
of $H_{N+1}(u)$ subject to $N$ constraints on the lower-order
Hamiltonians:
\begin{equation}
\label{constraints} H_n(u) = {\rm constant}, \qquad n = 1,...,N,
\end{equation}
so that the second variation of $\Lambda_N(u)$ is positive definite,
\begin{equation}
\label{positive} \frac{1}{2} \left( v, {\cal L}_N v \right) =
\lim_{\epsilon \to 0} \frac{\Lambda_N(U_N + \epsilon v) -
\Lambda_N(U_N)}{\epsilon^2} > 0, \quad v \in X_c({\Bbb R}) \cap
X_c'({\Bbb R}).
\end{equation}
Here ${\cal L}_N$ is the self-adjoint linearized operator of
$(2N)$-order with finite number of negative eigenvalues,
finite-dimensional kernel and positive continuous spectrum, bounded
away from zero (see \cite{MS93}), $X_c({\Bbb R})$ is the closed
orthogonal compliment of the kernel of ${\cal L}_N \partial_x$:
\begin{equation}
\label{constrained-space} X_c({\Bbb R}) = \left\{ v \in L^2({\Bbb
R}) : \; \left( v, \frac{\delta H_n}{\delta U_N} \right) = 0, \; n =
1,...,N \right\},
\end{equation}
and $X_c'({\Bbb R})$ is the closed orthogonal compliment of the
kernel of ${\cal L}_N$:
\begin{equation}
\label{constrained-space10} X_c'({\Bbb R}) = \left\{ w \in L^2({\Bbb
R}) : \; \left( w, \frac{\partial}{\partial x} \frac{\delta
H_n}{\delta U_N} \right) = 0, \; n = 1,...,N \right\}.
\end{equation}
The embedding $v \in X_c({\Bbb R})$ follows from the constraints
(\ref{constraints}), while the embedding $v \in X_c'({\Bbb R})$ is
set artificially, in order to remove the zero eigenvalues of ${\cal
L}_N$ and to ensure positivity of the energy quadratic form
(\ref{positive}). Due to positivity (\ref{positive}), the functional
$\Lambda(u)$ is convex at the point $u=U_N(x)$, such that Lyapunov
stability of $N$-solitons in energy space $H^N({\Bbb R})$ holds
\cite{MS93}.

The linearized operators ${\cal L}_1$ and ${\cal L}_2$ for
$1$-soliton and $2$-solitons are given explicitly as:
\begin{equation} \label{linear-1}
{\cal L}_1 = - \partial_x^2 - 6 U_1(x) + c_1,
\end{equation}
and
\begin{equation}
\label{linear-2} {\cal L}_2 = \partial_x^4 + 10 U_2(x)
\partial_x^2 + 10 U_2'(x) \partial_x + 10 U_2''(x) + 30 U_2^2(x) -
(c_1 + c_2) \left( \partial_x^2 + 6 U_2(x) \right) + c_1 c_2.
\end{equation}
The first two eigenfunctions of the kernel of ${\cal L}_N$ are:
\begin{equation}
v_1 = U_N'(x), \qquad v_2 = -U_N'''(x) - 6 U_N(x) U_N'(x).
\end{equation}
The first two eigenfunctions of the kernel of ${\cal L}_N
\partial_x$ are:
\begin{equation}
w_1 = U_N(x), \qquad w_2 = - U_N''(x) - 3 U_N^2(x).
\end{equation}
These explicit expressions illustrate the general construction of
the linearized operator ${\cal L}_N$ and the kernels of ${\cal L}_N$
and ${\cal L}_N \partial_x$.

\section{Spectral stability of $N$-solitons}

Time evolution of $N$-solitons (\ref{Nsoliton})--(\ref{theta}) is
defined in the mixed commuting flows of the KdV hierarchy, which are
generated by the Lyapunov functional (\ref{Lyapunov}):
\begin{equation}
\label{time-evolution} \frac{d u}{d t} = \frac{\partial u}{\partial
t_{N+1}} + \sum_{n=1}^N \nu_n \frac{\partial u}{\partial t_n} =
\frac{\partial}{\partial x} \frac{\delta \Lambda_N}{\delta u}.
\end{equation}
The family of $N$-solitons $u =
U_N(x;c_1,...,c_N;\delta_1,...,\delta_N)$ is a time-independent
space-decaying solution of the KdV-type evolution equation
(\ref{time-evolution}), where $\nu_1$,...,$\nu_N$ are elementary
symmetric functions of $c_1$,...,$c_N$. Linearization of the
evolution equation (\ref{time-evolution}) as $u(x;t) = U_N(x) +
V_N(x;t)$ and separation of variables as $V_N(x;t) = v(x) e^{\lambda
t}$ results in the spectral stability problem
\begin{equation}
\label{spectrum-N-sol}
\partial_x {\cal L}_N v = \lambda v
\end{equation}
where ${\cal L}_N$ is the same as in the energy quadratic form
(\ref{positive}). We assume that the family of $N$-solitons is
non-degenerate \cite{MS93} and characterize the kernel of ${\cal
L}_N$ in terms of the symmetries of the KdV hierarchy \cite{KT78}.

\begin{assumption}
\label{assumption-speeds} $N$-solitons
$U_N(x;c_1,...,c_N;\delta_1,...,\delta_N)$ have distinct positive
parameters $c_1$,...,$c_N$.
\end{assumption}

\begin{lemma}
\label{lemma-null} The kernel of ${\cal L}_N$ has a basis of $N$
linearly independent eigenfunctions $\{ v_n(x) \}_{n=1}^N$ in
$L^2({\Bbb R})$, where
\begin{equation}
\label{symmetry} v_n(x) = \frac{\partial}{\partial x} \frac{\delta
H_n}{\delta U_N}, \qquad n = 1,...,N.
\end{equation}
\end{lemma}

\begin{proof}
Derivatives of $U_N(x;c_1,...,c_N;\delta_1,...,\delta_N)$ with
respect to arbitrary parameters $\delta_1$,...,$\delta_N$ form a
basis of the kernel of ${\cal L}_N$, since they are linearly
independent and ${\cal L}_N$ is a self-adjoint operator of
$(2N)$-order. It follows from (\ref{theta}) that
\begin{equation}
\frac{\partial U_N}{\partial t_n} = \sum_{k=1}^N (-1)^{n-1} c_k^{n-1}
\frac{\partial U_N}{\partial \theta_k} = \sum_{k=1}^N (-1)^{n-1}
c_k^{n-1} \frac{\partial U_N}{\partial \delta_k}, \qquad n =
1,...,N.
\end{equation}
Since the Vandermonde determinant of $c_1$,...,$c_N$ is non-singular
under Assumption \ref{assumption-speeds}, the basis of $\left\{
\frac{\partial U_N}{\partial \delta_n} \right\}_{n=1}^N$ is
equivalent to the basis $\left\{ \frac{\partial U_N}{\partial t_n}
\right\}_{n=1}^N$. It follows from the KdV hierarchy
(\ref{KdVhierarchy}) that
\begin{equation}
\frac{\partial U_N}{\partial t_n} = \frac{\partial}{\partial x}
\frac{\delta H_n}{\delta U_N}, \qquad n = 1,...,N.
\end{equation}
See also \cite{KT78} and \cite[Lemma 3.4]{MS93} for alternative
proofs.
\end{proof}

\begin{lemma}
\label{lemma-generalized} The generalized kernel of $\partial_x
{\cal L}_N$ has a basis of $N$ linearly independent eigenfunctions
$\{ v_n^{(1)}(x) \}_{n=1}^N$ in $L^2({\Bbb R})$, where
\begin{equation}
\label{symmetry-generalized} v_n^{(1)}(x) = - \frac{\partial
U_N}{\partial \nu_n}, \;\; n = 1,...,N.
\end{equation}
\end{lemma}

\begin{proof}
The generalized kernel of $\partial_x {\cal L}_N$ is generated by
solutions of the inhomogeneous problem:
\begin{equation}
\label{generalized1}
\partial_x {\cal L}_N v_n^{(1)} = v_n(x).
\end{equation}
Integrating (\ref{generalized1}) in $x$ for $v_n^{(1)} \in L^2({\Bbb
R})$, we find that eigenfunctions $v_n^{(1)}(x)$ satisfy the
inhomogeneous equations,
\begin{equation} \label{generalized111}
{\cal L}_N v_n^{(1)} = \frac{\delta H_n}{\delta U_N}, \qquad
n = 1,...,N.
\end{equation}
It follows from the derivative of the variation equations
(\ref{Lyapunov-variation}) in $\nu_n$ that
\begin{equation}
\label{generalized2} {\cal L}_N \frac{\partial U_N}{\partial
\nu_n} = - \frac{\delta H_n}{\delta U_N}, \qquad n = 1,...,N.
\end{equation}
As a result, relations (\ref{symmetry-generalized}) hold.
\end{proof}

\begin{remark}
The symmetries $v_n(x)$ and $v_n^{(1)}(x)$ are shown to be expressed
by the squared eigenfunctions of the inverse spectral method
\cite{KN78,KT78}. The squared eigenfunctions form a basis for a class of
functions $u(x)$ satisfying the bounds:
\begin{equation}
\label{potential}
\int_{-\infty}^{\infty}(1+x^2)|u(x)|\, dx\, <\, \infty\,.
\end{equation}
\end{remark}

Define the solution surface as $\Lambda_s = \Lambda_N(U_N) =
\Lambda_s(\nu_1,...,\nu_N)$. The Hessian matrix ${\cal H}$ of
principal curvatures of the solution surface
$\Lambda_s(\nu_1,...,\nu_N)$ has the elements:
\begin{equation}
\label{hessian} {\cal H}_{n,m} = \frac{\partial^2
\Lambda_s}{\partial \nu_n \partial \nu_m} = \frac{\partial
H_n(U_N)}{\partial \nu_m} = \frac{\partial H_m(U_N)}{\partial
\nu_n}, \quad n,m = 1,...,N,
\end{equation}
where
\begin{equation}
\label{representation} \frac{\partial H_n(U_N)}{\partial \nu_m} =
\left( \frac{\partial U_N}{\partial \nu_m}, \frac{\delta
H_n}{\delta U_N} \right) = - \left( \frac{\partial U_N}{\partial
\nu_m}, {\cal L}_N \frac{\partial U_N}{\partial \nu_n} \right),
\quad n,m = 1,...,N.
\end{equation}
Let $z({\cal H})$ be the number of zero eigenvalues of ${\cal H}$.

\begin{lemma}
\label{lemma-second-kernel} The second generalized kernel of
$\partial_x {\cal L}_N$ is empty in $L^2({\Bbb R})$ if $z({\cal H})
= 0$.
\end{lemma}

\begin{proof}
The second generalized kernel is generated by solutions of the
inhomogeneous problem:
\begin{equation}
\label{generalized3}
\partial_x {\cal L}_N v^{(2)} = \sum_{n=1}^N a_n v_n^{(1)}(x),
\end{equation}
where $\{ a_n \}_{n=1}^N$ is a set of coefficients. Computing the
inner products of the inhomogeneous problem (\ref{generalized3})
with the set of eigenfunctions $\{\frac{\delta H_n}{\delta U_N}
\}_{n=1}^N$, we find that the solutions $v^{(2)} \in L^2({\Bbb R})$
may exist only if the Hessian matrix ${\cal H}$ in
(\ref{hessian})--(\ref{representation}) has a zero eigenvalue. They
do not exist if $z({\cal H}) = 0$.
\end{proof}

\begin{lemma}
\label{lemma-nullM} The kernel of ${\cal M}_N = -
\partial_x {\cal L}_N \partial_x$ has a basis of $N$
linearly independent eigenfunctions $\{ w_n(x) \}_{n=1}^N$ in
$L^2({\Bbb R})$, where
\begin{equation}
\label{symmetryM}
w_n(x) = \frac{\delta H_n}{\delta U_N}, \;\; n = 1,...,N,
\end{equation}
\end{lemma}

\begin{proof}
Integrating $\partial_x {\cal L}_N \partial_x w = 0$ in $x$, we have
${\cal L}_N w'(x) = 0$ for $w'(x) \in L^2({\Bbb R})$. Therefore, $w'
\in {\rm ker}({\cal L}_N)$, such that the eigenfunctions
(\ref{symmetryM}) follow from integration of eigenfunctions
(\ref{symmetry}).
\end{proof}

We now work with the coupled problem
\begin{equation}
\label{coupled-problem-N-sol} {\cal L}_N v = - \lambda w, \qquad
{\cal M}_N w = \lambda v,
\end{equation}
which is equivalent to the spectral problem (\ref{spectrum-N-sol})
when $v = - w'(x)$, subject to the constraint
(\ref{mass-constraint}). By Lemmas \ref{lemma-null} and
\ref{lemma-nullM}, definitions (\ref{Xc})--(\ref{Xc'}) of the
constrained spaces $X_c({\Bbb R})$ and $X_c'({\Bbb R})$ coincide
with the definitions
(\ref{constrained-space})--(\ref{constrained-space10}). It is
therefore clear that $v \in X_c(\mathbb{R})$ and $w \in
X_c'(\mathbb{R})$ for any solution $(v,w) \in
L^2(\mathbb{R},\mathbb{C}^2)$ of the coupled problem
(\ref{coupled-problem-N-sol}) with $\lambda \neq 0$.

Analysis of unstable eigenvalues in the coupled problem
(\ref{coupled-problem-N-sol}) is based on the fact that both
operators ${\cal L}_N$ and ${\cal M}_N$ have finitely many negative
eigenvalues in $L^2(\mathbb{R})$. The continuous spectrum of ${\cal
L}_N$ is positive and bounded away from zero by $c_0 = \prod_{n=1}^N
c_n > 0$, due to the exponential decay of potential functions and
the factorization relation \cite{MS93}:
\begin{equation}
\label{factorization} U_N(x) \equiv 0 : \quad {\cal L}_N = \left( -
\partial_x^2 + c_1 \right) ... \left( - \partial_x^2 + c_N \right).
\end{equation}
The continuous spectrum of ${\cal M}_N$ is non-negative, due to the
explicit relation: ${\cal M}_N = - \partial_x {\cal L}_N
\partial_x$. The kernel of ${\cal L}_N$ belongs to
$X_c({\Bbb R})$, due to commutativity of the Hamiltonians $H_n(u)$
of the KdV hierarchy (\ref{KdVhierarchy})--(\ref{HamiltonianH3}):
\begin{equation}
\left( \frac{\delta H_n}{\delta U_N}, \frac{\partial}{\partial x}
\frac{\delta H_m}{\delta U_N} \right) = 0, \quad n,m = 1,...,N.
\end{equation}
Equivalently, the kernel of ${\cal M}_N$ belongs to $X_c'({\Bbb
R})$. These facts enable us to consider the number of negative
eigenvalues of ${\cal L}_N$ and ${\cal M}_N$ in $X_c(\mathbb{R})$
and $X_c'(\mathbb{R})$, respectively, and to relate the absence of
negative eigenvalues to spectral stability of $N$-solitons in the
spectral problem (\ref{spectrum-N-sol}).

\begin{proposition}
\label{lemma-negative} Let $n({\cal L}_N)$ be the number of negative
eigenvalues of ${\cal L}_N$ in $L^2({\Bbb R})$ and $p({\cal H})$ be
the number of positive eigenvalues of ${\cal H}$. The numbers
$\#_{<0}({\cal L}_N)$ and $\#_{<0}({\cal M}_N)$ of negative
eigenvalues of ${\cal L}_N$ in $X_c({\Bbb R})$ and of ${\cal M}_N$
in $X_c'({\Bbb R})$, respectively, are given by:
$$
\#_{<0}({\cal L}_N) = \#_{<0}({\cal M}_N) = n({\cal L}_N) -
p({\cal H}).
$$
\end{proposition}

\begin{proof}
It follows from \cite[Theorem 3.1]{GSS90} and \cite[Lemma
2.1]{MS93} that $\#_{<0}({\cal L}_N) = n({\cal L}_N) - p({\cal
H})$. Equivalently, the same result can be proved by minimization
of quadratic forms with Lagrange multipliers, see
\cite{P03,CPV03}. In order to prove the relation $\#_{<0}({\cal
M}_N) = \#_{<0}({\cal L}_N)$, we integrate the quadratic form
$\left( w,{\cal M}_N w \right)$, $w \in X_c'({\Bbb R})$ by parts:
\begin{equation}
\label{diagonal-form} \left( w, {\cal M}_N w \right) = - \left( w,
\partial_x {\cal L}_N \partial_x w \right) = \left( w', {\cal
L}_N w' \right).
\end{equation}
By integrating the constraints in the inner products
(\ref{constrained-space10}) by parts, we confirm that if $w \in
X_c'(\mathbb{R})$, then $w' \in X_c({\Bbb R})$, such that the
quadratic form $\left( w', {\cal L}_N w' \right)$ is defined in $w'
\in X_c({\Bbb R})$. Finally, it follows from (\ref{symmetry}) and
(\ref{symmetryM}) after integration by parts that
\begin{equation}
\left( v_n,{\cal M}_N^{-1} v_m \right) = \left( w_n, {\cal L}_N^{-1}
w_m \right), \qquad n,m = 1,...,N,
\end{equation}
where the functions ${\cal M}_N^{-1} v_m$ are bounded due to the
fact that $v_m(x)$  satisfy the constraint (\ref{mass-constraint})
for any $m$. Applying Lemma 3.4 from \cite{CPV03}, we have the
relation $\#_{<0}({\cal M}_N) = n({\cal L}_N) - p({\cal H}) =
\#_{<0}({\cal L}_N)$.
\end{proof}

\begin{proposition}
\label{proposition-bounds} Let $p({\cal H}) = n({\cal L}_N)$. The
coupled problem (\ref{coupled-problem}) has no eigenvalues $\lambda
\in \mathbb{C}$ with ${\rm Re}(\lambda) > 0$ and $(v,w) \in
L^2(\mathbb{R},\mathbb{C}^2)$.
\end{proposition}

\begin{proof}
When $p({\cal H}) = n({\cal L}_N)$, we have by Proposition
\ref{lemma-negative}:
\begin{eqnarray}
\forall v \in X_c({\Bbb R}) & : & \quad \left( v, {\cal L}_N v
\right) \geq 0, \\ \forall w \in X_c'({\Bbb R}) & : & \left( w,
{\cal M}_N w \right) \geq 0,
\end{eqnarray}
where the zero values are achieved if and only if $v \in {\rm
ker}({\cal L}_N)$ in $X_c(\mathbb{R})$ and $w \in {\rm ker}({\cal
M}_N)$ in $X_c'(\mathbb{R})$. We assume that there exists an
eigenvalue $\lambda_0 \in \mathbb{C}$ with ${\rm Re}(\lambda_0) > 0$
and $(v_0,w_0) \in L^2(\mathbb{R},\mathbb{C}^2)$ and show the
contradiction. If the eigenvalue $\lambda_0$ would exist, then the
coupled system (\ref{coupled-problem-N-sol}) would result in the
identity:
\begin{equation}
\lambda_0 (\bar{w}_0,{\cal M}_N w_0) + \bar{\lambda}_0
(\bar{v}_0,{\cal L}_N v_0) = 0,
\end{equation}
where the real-valued inner products are used for clarity of
notations. Since ${\rm Re}(\lambda_0) > 0$ and quadratic forms
associated with ${\cal L}_N$ and ${\cal M}_N$ are real-valued, we
have
$$
(\bar{w}_0,{\cal M}_N w_0) = - (\bar{v}_0,{\cal L}_N v_0),
$$
which is the contradiction, since both quadratic forms are
non-negative for $v_0 \in X_c(\mathbb{R})$ and $w_0 \in
X_c'(\mathbb{R})$ and non-zero for $v_0 \notin {\rm ker}({\cal
L}_N)$ and $w_0 \notin {\rm ker}({\cal M}_N)$ (when $\lambda_0 \neq
0$).
\end{proof}

\noindent {\bf Spectral Stability Theorem:} {\em Let $N$-solitons
$U_N(x)$ of the KdV hierarchy
(\ref{KdVhierarchy})--(\ref{HamiltonianH3}) satisfy Assumption
\ref{assumption-speeds}. Then, $N$-solitons are weakly spectrally
stable such that the spectral problem (\ref{spectrum-N-sol}) has no
eigenvalues $\lambda \in \mathbb{C}$ with ${\rm Re}(\lambda) > 0$
and $v \in L^2(\mathbb{R},\mathbb{C})$.}

\begin{proof}
We show that conditions of Proposition \ref{proposition-bounds} are
satisfied for $N$-solitons of the KdV hierarchy
(\ref{KdVhierarchy})--(\ref{HamiltonianH3}). It was proved in
\cite[Lemma 3.5]{MS93} and in \cite[Lemma 3.6]{MS93} that
\begin{equation}
n({\cal L}_N) = \left[ \frac{N+1}{2} \right], \qquad p({\cal H}) =
\left[ \frac{N+1}{2} \right],
\end{equation}
where $[z]$ is the integer part of $z$. Therefore, the result holds
by Proposition \ref{proposition-bounds}.
\end{proof}

\begin{remark}
It was also proved in \cite[Lemma 3.7]{MS93} that $z({\cal H}) = 0$.
Therefore, the second generalized kernel of $\partial_x {\cal L}_N$
is empty by Lemma \ref{lemma-second-kernel}, such that the zero
eigenvalue $\lambda = 0$ in the spectral problem (\ref{spectrum}) is
controlled by the symmetries of the KdV hierarchy.
\end{remark}

\section{Time evolution of $N$-solitons in the KdV hierarchy}

According to (\ref{positive}), the Lyapunov functional
$\Lambda_N(u)$ defined in (\ref{Lyapunov}) is convex at the point
$u=U_N(x)$, where $U_N(x)$ is $N$-solitons of the KdV hierarchy
(\ref{KdVhierarchy})--(\ref{HamiltonianH3}). We give a direct proof
of convexity based on the inverse spectral method \cite{Newell}.
Moreover, we prove uniqueness of $N$-solitons as minimizers of the
constrained variational problem
(\ref{Lyapunov-variation})--(\ref{constraints}). Using the
asymptotic results from \cite{AK82}, we formulate the Time Evolution
Theorem for $N$-solitons in the energy space $H^N(\mathbb{R})$,
which follows from Lyapunov stability of $N$-solitons.

These results improve the Lyapunov Stability Theorem of \cite{MS93},
where the time evolution of parameters $\theta_n(t)$ in the solution
$U_N(x)$ remains undefined and uniqueness of minimizers in the
constrained variational problem
(\ref{Lyapunov-variation})--(\ref{constraints}) remains open. The
unique characterization of the time evolution of parameters of $N$
solitons is equivalent to the asymptotic stability of $N$ solitons
in the mixed commuting time flows (\ref{time-evolution}). We note
that asymptotic stability of 1-soliton in the KdV equation was
proved in \cite{PW94} with exponentially weighted spaces (see also
\cite{Ben,B75} for pioneer papers), while asymptotic stability of
$N$-solitons in the energy space $H^1({\Bbb R})$ was recently proved
in \cite[Corollary 1]{MMT02} with different analysis of energy
functionals.

Let us recall some necessary formulae obtained in the inverse
spectral method \cite{Newell}. The method is based on the
isomorphism between the potential $u(x,t)$ of the Schr\"odinger
equation and the scattering data $S(t)$ (see, e.g., \cite{Newell}).
Consider the scattering problem,
\begin{equation}
\label{schroedinger} \left(\partial_x^2 + u(x,t)+k^2\right)\psi=0\,,
\end{equation}
with the boundary conditions,
\begin{equation}
\label{boundary} \psi(x,t;k)\longrightarrow\left\{
\begin{array}{ccccc}
e^{-ikx}       &  {\rm as}  &  x\to -\infty \\
a(k)e^{-ikx}+b(t;k)e^{ikx} & {\rm as} & x\to\infty\\
\end{array}
\right..
\end{equation}
Assuming that the potential $u(x,t)$ is exponentially decaying in
$x$, it is proved that the function $a(k)$ is analytic on the upper
half plane of $k\in{\mathbb C}$, and the bound states are defined by
the zeros of $a(k)$ at $k=i\kappa_j,\,\kappa_j>0$, such that
\begin{equation}
a(i\kappa_j)=0\, \quad j=1,\ldots, N.
\end{equation}
The scattering data is then defined by
\begin{equation}
\label{scattering} S(t)=\left[\,\{\kappa_j, C_j(t)\}_{j=1}^N\,,\,\,
\{ r(t;k) \}_{k\in{\mathbb R}} \,\right]\,,
\end{equation}
where $r(t;k) = b(t;k)/a(k)$ is the reflection coefficient, and
$C_j(t)$ is the normalization constant (see \cite{Newell} for
details). The parameter $c_j$ in the $N$-solitons
(\ref{Nsoliton})--(\ref{theta}) is given by $c_j = 4\kappa_j^2$ for
all $j$. We assume that the parameters $\{ \kappa_j \}_{j=1}^N$ are
ordered as
\begin{equation}
\label{order} \kappa_1 > \kappa_2 > \cdots > \kappa_N>0\,.
\end{equation}
By inverse scattering, the potential $u(x,t)$ can be expressed as
\begin{equation}
\label{usolution}
u(x,t)=4\sum_{j=1}^{N}\kappa_jC_j(t)\psi^2(x,t;i\kappa_j)
+\frac{2i}{\pi} \int_{-\infty}^{\infty}kr(t;k)\psi^2(x,t;k)\,dk\,,
\end{equation}
which shows that $u(x,t)$ consists of $N$-solitons and the
radiation. With the formula (\ref{usolution}), the Hamiltonian $H_n$
can be expressed in terms of the scattering data \cite{Newell},
\begin{equation}
\label{scathamilton}
H_n(u)=(-1)^{n+1}2^{2n+1}\left(\frac{1}{2n+1}\sum_{j=1}^N\kappa_j^{2n+1}
-(-1)^{n} R_n\right),
\end{equation}
where $R_n$ is the radiation part, given by
\begin{equation}
\label{radiation} R_n=-\frac{1}{2\pi}\int_{0}^{\infty}
k^{2n}\ln(1-|r(t;k)|^2)\,dk\,\ge 0, \quad n = 1,2,...
\end{equation}
We show that the alternate sign $(-1)^n$ in front of $R_n$ in the
representation (\ref{scathamilton}) plays a crucial rule for the
convexity.

\begin{proposition}
\label{proposition-uniqueness} The $N$-soliton $U_N(x)$ is uniquely
determined in the variational problem (\ref{Lyapunov-variation}) by
the constraints (\ref{constraints}) except the phase parameters
$\delta_1,\ldots,\delta_N$. The Lyapunov functional $\Lambda_N(u)$
is convex at $u = U_N(x)$.
\end{proposition}

\begin{proof}
Let us consider the variation of $H_n(u)$ at $u = U_N(x)$, which we
denote as $\Delta H_n(U_N)$. The constraints (\ref{constraints})
imply that $\Delta H_n(U_N) = 0$, such that
\begin{equation}
\label{constraint} \sum_{j=1}^{N}\kappa_j^{2n}\Delta\kappa_j -
(-1)^n \Delta R_n = 0, \quad n=1,\ldots,N,
\end{equation}
where $\Delta \kappa_j$ is the variation of $\kappa_j$, and $\Delta
R_n$ is the variation of $R_n$. Since $R_n = 0$ at $u = U_N(x)$, we
understand from the explicit formula (\ref{radiation}) that all
$\Delta R_n \geq 0$. It is clear from the system (\ref{constraint})
that the variations of $\kappa_j$ are balanced with the variations
of $R_n$, so that the Hamiltonian $H_n$ remains constant. The
variations $\Delta U_N$ belongs then to the constrained space
$X_c({\Bbb R})$, defined in (\ref{constrained-space}). We also note
that the number of solitons may change under the variation of
$H_n(u)$, but this does not appear in the constraints
(\ref{constraint}), since $\kappa_{j}=0$ for $j>N$. We need to prove
the convexity of the functional $\Lambda_N(u)$ at $u=U_N(x)$, which
implies that
\begin{equation}
\label{convex} \Delta H_{N+1}(U_N) > 0.
\end{equation}
>From the system (\ref{constraint}), the variation $\Delta
H_{N+1}(U_N)$ can be expressed as
\begin{equation}
\label{deltaH}
\begin{array}{lllllll}
\Delta H_{N+1}(U_N) & = & \displaystyle{(-1)^{N}2^{2N+3}
\left(\sum_{j=1}^{N} \kappa_j^{2N+2} \Delta \kappa_j
+ (-1)^{N}\Delta R_{N+1}\right)} \\
& = & \displaystyle{\frac{(-1)^N2^{2N+3}}{D} \left(
\sum_{l=1}^{N}(-1)^l D_l \Delta R_l + (-1)^N \Delta R_{N+1}
\right)},
\end{array}
\end{equation}
where $D$ is the determinant of the $N\times N$ matrix
$K_N:=\{(\kappa_i^{2j})\,: \, {1\le i,j\le N}\}$ and $D_l$ is the
determinant of the matrix $K_N$ after replacing the $l$-th row by
$\{(\kappa_i^{2N+2})\,: \, 1\le i \le N\}$. It is clear that $D$ is
the Vandermonde determinant, computed as
\begin{equation}
D = \prod_{1\le i<j\le N}(\kappa_i^2-\kappa_j^2)\,,
\end{equation}
so that $D > 0$ from the ordering (\ref{order}). We further notice
that $D_l$ can be written as
\begin{equation}
D_l = (-1)^{N-l} D_l^{N+1},
\end{equation}
where $D_l^{N+1}$ is the determinant of the $(N+1)\times(N+1)$
matrix $K_{N+1}=\{(\kappa_i^{2j})\,:\, 1\le i,j\le N+1\}$ after
removing the $l$-th row and the $(N+1)$-th column. Thus we have
\begin{equation}
\Delta H_{N+1}(U_N)=\frac{2^{2N+2}}{D} \left(\sum_{l=1}^{N}
D^{N+1}_l \Delta R_l + \Delta R_{N+1}\right)\,.
\end{equation}
Since all $\Delta R_n \geq 0$, we only need to show that
$D_l^{N+1}>0$ for the convexity. This follows from the expression,
\begin{equation}
D_{l}^{N+1} = \sigma_N \sigma_{N-l+1} D,
\end{equation}
where $\sigma_k$ is the symmetric polynomial of
$(\kappa_1^{2},\ldots,\kappa_N^2)$ given by
\begin{equation}
\sigma_k = \sum_{1\le i_1<\cdots<i_k\le
N}\kappa_{i_1}^2\cdots\kappa_{i_k}^2\, >0.
\end{equation}
As a result, $D_l^{N+1} > 0$ and the convexity (\ref{convex}) holds
if there exists at least one $\Delta R_n \neq 0$. (Note that if
$\Delta R_n\ne 0$ for some $n$, then it is not zero for any $n$.)
Otherwise, i.e. when all $\Delta R_n = 0$, all $\Delta \kappa_n = 0$
from the system (\ref{constraint}), and the variations of $U_N(x)$
simply translate the solution $U_N(x)$ along its phase parameters
$\delta_1$,...,$\delta_N$. When the variation $\Delta U_N \in
X_c({\Bbb R})$ belongs also to the constrained space $X_c'({\Bbb
R})$, defined in (\ref{constrained-space10}), the latter case is
excluded. Thus the minimal point of $H_{N+1}$ is given by the
$N$-soliton $u = U_N(x)$, that is, all $R_n=0$. Uniqueness of the
minimizer $U_N(x)$ follows from the geometry. Each constraint
$H_n(u) = {\rm constant}$ gives a hypersurface in
$(\kappa_1,\kappa_2,...,\kappa_N) \in {\mathbb R}_+^N$. If all those
hypersurfaces intersect transversally with nonempty intersection, we
have $N!$ points of the intersection. Then the ordering
$\kappa_1>\cdots>\kappa_N>0$ chooses a unique point of the
intersection.
\end{proof}

\medskip
\noindent {\bf Time Evolution Theorem:} {\it For all $\epsilon>0$,
there exists $\delta>0$ such that if $\Vert
u(x,0)-U_N(x)\Vert_{H^N({\mathbb R})}\le \delta$, then there exist
$\theta_j^{\pm}(t)=c_jt+\delta_j^{\pm}$, where $t\equiv t_2$, such
that}
\begin{equation}
\Vert  u(x,t)-U_N(x-\theta_1^{\pm}(t),\ldots, x-\theta_N^{\pm}(t))
\Vert_{L^2(|x|>c_N|t|)}\le \epsilon,\quad {\rm as}\quad
t\to\pm\infty.
\end{equation}

\begin{proof}
The proof relies on Proposition \ref{proposition-uniqueness} and
formalism above. The time evolution of $N$-solitons has been
obtained in \cite{AK82} by the inverse spectral method, see Eqs.
(2.15a) and (2.15b) in \cite{AK82}, where $\kappa_N t$ should be
read as $4 \kappa_N^2 t = c_N t$. We only note that the soliton
speeds $c_j = 4\kappa_j^2$ and soliton phases $\delta_j$ must be
obtained by the scattering problem (\ref{schroedinger}) for the
perturbed potential $u = U_N + \Delta U_N$.
\end{proof}

\section{Summary}

We have studied the spectral stability and time evolution of
$N$-solitons in the mixed commuting flows of the integrable KdV
hierarchy. We have proved that $N$-solitons of the KdV hierarchy are
spectrally stable with respect to the time evolution. The analysis
extend the recent results of \cite{CPV03,KKS03,P03} to the KdV-type
evolution equation (\ref{generalKdV}). In particular, the proof of
spectral stability is related to the absence of negative eigenvalues
of linearized energy in constrained function spaces. We have also
characterized time evolution of $N$-solitons and proved that the
$N$-solitons are uniquely determined in a constrained variational
problem. Further development of the spectral stability analysis to
non-integrable KdV-type evolution equations, e.g. to the fifth-order
KdV model, will be reported elsewhere.

{\bf Acknowledgements:} Y.K. is partially supported by the NSF grant
DMS-0404931. D.P. is partially supported by the NSERC Discovery
grant.

\end{document}